\documentstyle[alife7,epsf]{article}

\title{An Ecolab Perspective on the Bedau Evolutionary Statistics}

\author{
Russell K. Standish\\
High Performance Computing Support  Unit\\
University of New South Wales\\
Sydney, 2052\\
Australia\\
R.Standish@unsw.edu.au\\
http://parallel.hpc.unsw.edu.au/rks
}

\newcommand{\cit}[2]{#2}

\newcommand{\br}{\mbox{\boldmath{$r$}}}          
\newcommand{\bbeta}{\mbox{\boldmath{$\beta$}}}   
\newcommand{\bgamma}{\mbox{\boldmath{$\gamma$}}} 
\newcommand{\bmu}{\mbox{\boldmath{$\mu$}}}       
\newcommand{\bn}{\mbox{\boldmath{$n$}}}          
\newcommand{\nsp}{\mbox{$n_{\rm sp}$}}           
\newcommand{\htmladdnormallinkfoot}[2]{#1\footnote{#2}}
\newcommand{\Acum}{\bar{A}_{\mbox{\rm\scriptsize cum}}}
\newcommand{\Anew}{A_{\mbox{\rm\scriptsize new}}}

\begin{document}

\maketitle

\begin{abstract}
  At Alife VI, Mark Bedau proposed some evolutionary statistics as a
  means of classifying different evolutionary systems.  Ecolab, whilst
  not an artificial life system, is a model of an evolving ecology
  that has advantages of mathematical tractability and computational
  simplicity. The Bedau statistics are well defined for Ecolab, and
  this paper reports statistics measured for typical Ecolab runs, as a
  function of mutation rate. The behaviour ranges from class 1 (when
  mutation is switched off), through class 3 at intermediate mutation
  rates (corresponding to scale free dynamics) to class 2 at high
  mutation rates. The class 3/class 2 transition corresponds to an
  error threshold. Class 4 behaviour, which is typified by the
  Biosphere, is characterised by unbounded growth in diversity. It
  turns out that Ecolab is governed by an inverse relationship between
  diversity and connectivity, which also seems likely of the
  Biosphere. In Ecolab, the mutation operator is conservative with
  respect to connectivity, which explains the boundedness of
  diversity. The only way to get class 4 behaviour in Ecolab is to
  develop an evolutionary dynamics that reduces connectivity of time.
\end{abstract}

\section{Introduction}

At Alife VI, Mark Bedau proposed some evolutionary
statistics\cit{Bedau-etal98}{ (Bedau {\em et al.}, 1998)} as a means
of classifying different evolutionary systems. The intent here is to
find a general scheme analogous to Wolfram's\cit{Wolfram84}{ (1984)}
classification scheme of cellular automata. Three statistics are
proposed:
\begin{description}
\item[Diversity ($D$):] The number of species or components in the system
\item[Mean Cumulative Evolutionary Activity ($\Acum$):]
  Activity of a species is defined as the population count of that
  species, the vector \bn\ in Ecolab terms. Evolutionary activity
  subtracts from this the neutral or nonadaptive part. This is
  achieved by running a {\em neutral shadow model}, that is identical
  with the original model, except that natural selection must be
  ``turned off''. Finally, this activity is accumulated over the
  lifetime of the species, and then averaged over all species.
\item[New Evolutionary Activity ($\Anew$):] This corresponds
  the the number of new species crossing a threshold, divided by the
  diversity. 
\end{description}

Bedau describes four classes of evolutionary behaviour, as in the
following table:

\noindent{\small
\newlength{\descwidth}
\settowidth{\descwidth}{Description}
\begin{tabular}{|c|c|c|c|p{\descwidth}|}
\hline
Class & $D(t)$ & $\Acum(t)$ & $\Anew(t)$ &
Description \\
\hline\hline
1 & bounded & zero & zero & none \\\hline
2 & bounded & unbounded & none & unbounded, uncreative \\\hline
3 & bounded & bounded & positive & bounded, creative \\\hline
4 & unbounded & positive & positive & unbounded, creative \\\hline
\end{tabular}
}

Note that in \cit{Bedau-etal98}{Bedau {\em et al.} (1998)}, only 3
classes are mentioned --- class 2 was added later in his presentation
at Alife VI. Bedau has applied his statistics to a number of artificial
life models, including Echo\cit{Holland95}{ (Holland, 1995)} and
Tierra\cit{Ray91}{ (Ray,1991)}, none of which exhibit class 4
behaviour. By contrast, the same statistics applied to the fossil
record (at least for the Phanerozoic --- the period of time since the
appearance of multicellular life in the Cambrian) --- show a strong
class 4 behaviour. Further, Bedau speculates that the global economy
and internet traffic are also class 4, particularly as they show
strong growth over a significant period of time. Since no artificial
life systems to date appear to show class 4 behaviour, the gauntlet
has been laid down to discover such a system to work out whether this
classification difference is fundamental or not.

Ecolab\cit{Standish94}{ (Standish, 1994)}, whilst not an artificial
life system, is a model of an evolving ecology that has advantages of
mathematical tractability and computational simplicity. It lies in
between the extremely simplistic models of (for example)
\cit{Bak-Sneppen93}{Bak and Sneppen (1993)} or \cit{Newman97b}{ Newman
  (1997)} and artificial life models of evolution such as {\em Tierra}
or {\em Avida}. One of its key characteristics is that its dynamics
are defined by the ecological interactions between the species, rather
than ad hoc exogenous dynamics. The Bedau statistics are well defined
for it, so it is interesting to see what class behaviour Ecolab has.
Furthermore, an Ecolab-like model is possible for all artificial life
systems (valid in a continuum limit). For example, the equations of
motion for Tierra are given in Standish\cit{Standish97b}{(Standish,
  1997)}.

\section{Ecolab}\label{Ecolab-sect}

The Ecolab model (as opposed to the Ecolab simulation system) is based
on an evolving Lotka-Volterra ecology.  The defining equation is given
by:
\begin{equation}\label{Ecolab}
\dot{\bn} = \br*\bn + \bn*\bbeta\bn + \bgamma*\nabla^2\bn + \bmu(\br*\bn)
\end{equation}
where \bn{} is the species density, \br{} the effective reproduction
rate (difference between the intrinsic birth and death rates in the
absence of competition), \bbeta{} the matrix of interaction terms
between species, \bgamma{} the migration rate and \bmu{} the mutation
operator. All of these quantities (apart from \bbeta, which is a
matrix) are vectors of length \nsp, the number of species in the
ecology. The operator $*$ denotes elementwise multiplication. The
mutation operator returns a vector of dimensionality greater than
\nsp, with the first \nsp{} elements set to zero --- in effect
expanding the dimensionality of the space, a key feature of this
system. For a more detailed exposition of the various properties of
the model, in particular, the precise form of the mutation operator,
the reader is referred to the previous published papers, as well as
the Ecolab Technical Report\cit{Ecolab-Tech-Report}{}, which are all
available from the \htmladdnormallinkfoot{Ecolab Web
Site}{http://parallel.hpc.unsw.edu.au/rks/ecolab.html}.

For the purposes of this paper, it is worthwhile expounding a little
on the properties of the mutation operator. It models point mutations
in particular (other mutation types, such as recombination are simply
not modeled within Ecolab). Point mutations in genotype space, which
satisfy Poisson statistics, give rise random mutations, with locality,
in phenotype space. Since the only phenotypic properties of interest
to the model are the parameters \br, \bbeta\ and \bgamma\, the
parameters are mutated according to a normal or lognormal distribution
(according as the parameters are reals or positive (or negative)
respectively), using a sample from the Poisson distribution for the
width. The two parameters governing mutation (width of the Poisson
distribution, and the rate at which mutations are attempted) are
related via a simple proportional factor (called the ``species radius
(or separation)'') that is kept constant throughout the simulations
reported here. Each species has its own mutation rate --- given as a
vector \bmu. 

Each of these phenotypic parameters are initialised from a uniform
distribution. The relevant input parameters for a run are then maximum
and minimum values for each of \br, the diagonal of \bbeta, the
offdiagonal of \bbeta, \bmu, \bgamma{} and the species radiua $\rho$.
The complete system may be scaled in the time dimension, fixed by what
value is chosen for the timestep. In this case, $\max_ir_i=0.1$, so
one timestep corresponds to about a 14th of the doubling time of the
fastest reproducing organism in the ecology. This is a compromise
between continuity of the simulation and computational expense. The
ratio $\frac{\max_ir_i}{\max_i\beta_{ii}}$ roughly corresponds to the
carrying capacity of the ecology. This is chosen to about 100 so that
behaviour near the equilibrium is reasonably continuous rather than
stochastic. The ratio of offdiagonal to diagonal terms relates to how
negative definite \bbeta is. Since mutations tends to drive the matrix
away from being negative definite (system stability), the maximum of
the offdiagonal terms is chosen to make the initial system marginally
unstable. The species radius $\rho=0.1$ was chosen empirically to make
new species {\em phenotypically} distinct from its parent species.

Having fixed the other parameters according to the above criteria, the
remaining degrees of freedom are \bmu{} and \bgamma. In this paper, we
vary the maximum mutation rate in different simulations, but keep the
distribution of migration rates fixed.

One other feature worth noting is that the mutation operator will also
randomly add or drop connections between species, according to an
exponential distribution. Thus, the mutation operator is in fact
highly conservative --- with the lognormally mutated parameters capped
(in the case of \bmu{} and \bgamma) or restricted by the requirements of
boundedness (diagonal components of
\bbeta)\cit{Standish98b,Ecolab-Tech-Report}{(Standish, 1998; Ecolab
  Technical Report)}.

\section{Neutral Shadow Model}

An important feature for improving the accuracy of the evolutionary
statistics is the use of a {\em neutral shadow model}. This model
should be as similar as possible to the original model, but with all
selection turned off. In the case of Ecolab, this is accomplished by
running a shadow population density vector \bn', and when \bn\ is
updated, the shadow vector is updated by a random permutation of the
updates. Thus each shadow species behaves in the long run like an
average species. Activity is also tracked at the same time, with the
activity vector being updated by the difference between the population
density and the shadow population density, provided that difference is
positive.

The new activity statistic $\Anew$ is computed by summing the number
of species that have crossed a threshold. In \cit{Bedau-etal98}{(Bedau
  {\em et al.}, 1998)}, this threshold is determined by plotting the
activity distributions for both the original and the shadow model, and
taking the cross-over point as the threshold. This turned out to be 50
individuals, rather than the arbitrary 10 individuals used in other
Ecolab studies. In fact the two distributions are nearly equal over
the range 10--50, but if an activity is above 50, then it is highly
likely to be due to adaptive behaviour.

\begin{figure}[t]
\begin{center}
\mbox{}\epsfxsize=.6\textwidth\epsfbox{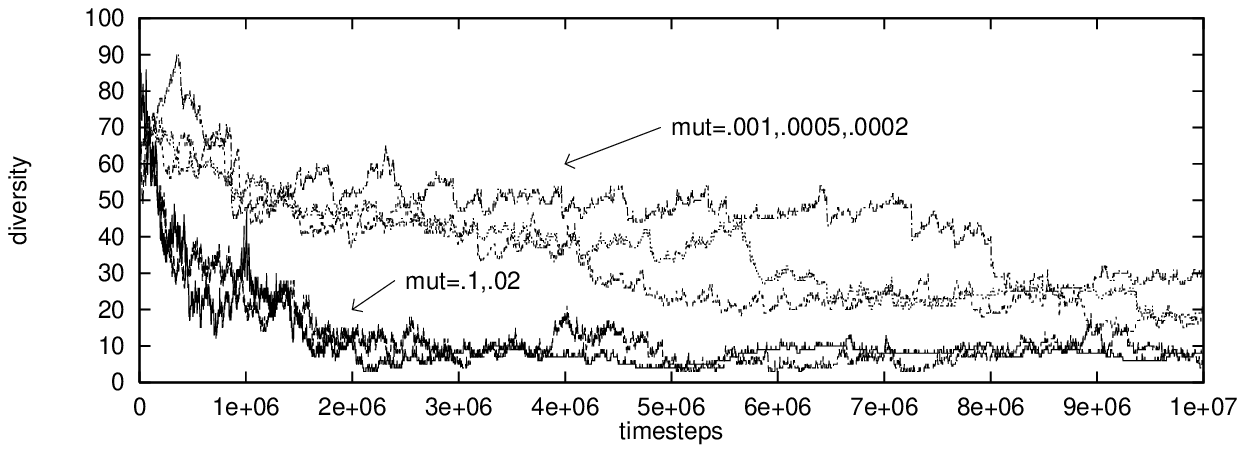}\\
\mbox{}\epsfxsize=.6\textwidth\epsfbox{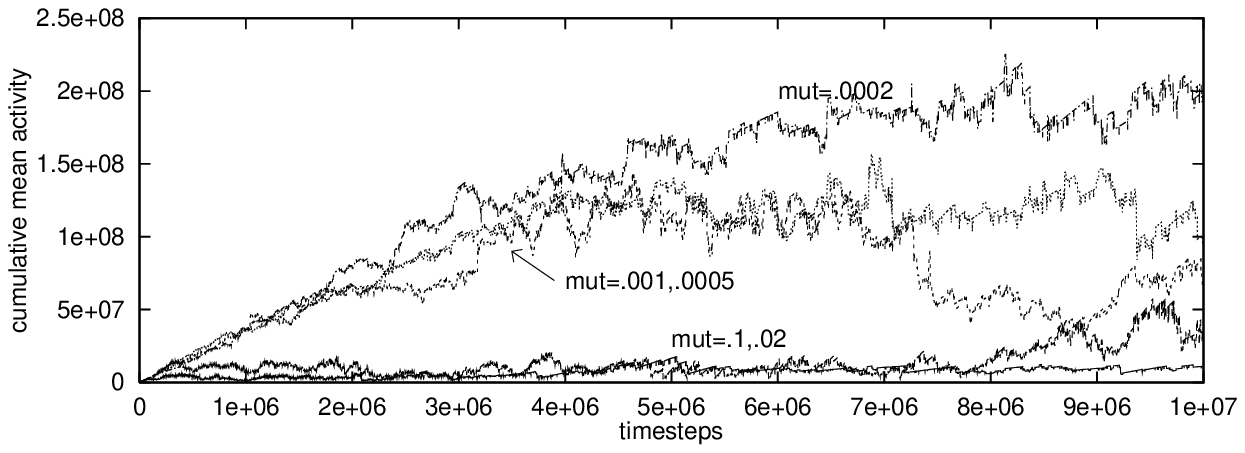}\\
\mbox{}\epsfxsize=.6\textwidth\epsfbox{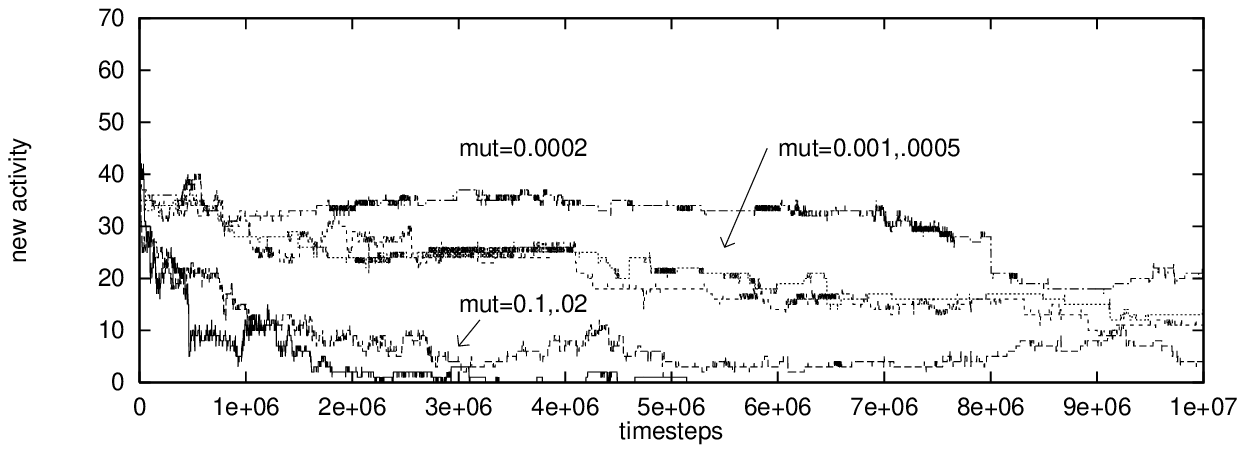}\\
\end{center}
\caption{A typical run for panmictic Ecolab at varying mutation rates, showing
  the Bedau statistics: diversity, cumulative mean activity and new
  activity}
\label{single cell}
\end{figure}

\section{Behaviour of Ecolab}

Figure \ref{single cell} shows the Bedau statistics for typical Ecolab
runs (panmictic, or spatially independent case), as a function of
mutation rate. When the mutation rate is too low, class two behaviour
is seen. Diversity remains constant, and activity grows unbounded as
the system rapidly sheds unviable organisms and tends to a stable
ecology. Conversely, for high levels of mutation, class one behaviour
is seen. There is a constant churn of organisms, that do not have any
chance to generate activity. For intermediate levels of mutation, an
interesting situation arises. Here, the number of mutant organisms
that successfully invade the ecosystem roughly balances the number
lost through extinction\cit{Standish98a}{ (Standish, 1998)}. Scale free
behaviour is observed in a number of statistics, such as the
distribution of species lifetime.  These same 3 states of behaviour
have been observed in Avida\cit{Adami-etal98a}{ (Adami {\em et al.},
  1998)}.

\hangindent=2cm
\hangafter=4
The code used for this simulation is available from the Ecolab web
site as version 3.3 of the software. The model including the neutral
shadow model is defined in {\tt shadow.cc}, and a sample experimental
script given as {\tt bedau.tcl}. The only parameters varied are the
spatial dimensions and {\tt mutation(random,maxval)}.



\hangindent=2cm
\hangafter=0
The evolutionary statistics were also collected for a spatially
dependent Ecolab, however due to some implementation difficulties, run
lengths exceeding $1\times10^6$ timesteps have not been achieved prior
to this paper's deadline. Broadly speaking, though, the same behaviour
is seen as the panmictic case, although there is a period of diversity
growth in the early period prior to settling on a higher level of
diversity than the panmictic case. 

\hangindent=2cm
\hangafter=-14
This can be understood by considering two extremes of spatially
dependent Ecolab models, namely zero migration and infinite migration.
Infinite migration effectively corresponds to the panmictic case
again, whereas zero migration corresponds to a number of cells,
independent of each other, each running the panmictic model. So we
would expect in the case of zero migration, the diversity (in the long
run) should be proportional to the number of cells (or the total
area). The in between case of finite nonzero migration should also
show an increase in diversity with area, due to partial independence
of each cell, but the increase should be sublinear, as migration
causes some species to be identified between cells. Island
Biogeography\cit{MacArthur-Wilson67}{ (MacArthur and Wilson, 1967)}
theory postulates that the relationship is $D\propto A^{-s}$ for some
coefficient $s$, which presumably must depend in some fashion on the
migration rates, but is generally in the range 0.2--0.35 for most
empirical studies.

\section{May's Stability Criterion}

May\cit{May72}{ (1972)} proposed that random Lotka-Volterra webs would be
unstable if
\begin{equation}\label{May's criterion}
\nsp<\frac{1}{s^2C}
\end{equation}
where $C$ is the connectivity,\index{connectivity} defined as the
proportion of nonzero elements in \bbeta, and $s$ is the interaction
strength, defined as the standard deviation of the offdiagonal terms
of \bbeta, divided by the average of the diagonal terms. Cohen and
Newman\cit{Cohen-Newman85}{ (1985)} showed that May's criteria does not hold
for Lotka-Volterra systems in general, only a smaller class related to
the models May studied. However, the inverse relationship between
species number and connectivity does appear to
hold\cit{Pimm82,Cohen-Newman88,Cohen-etal90}{ (Pimm, 1982;Cohen and
Newman, 1988;Cohen {\em et al.}, 1990)}.

Stability is not a relevant property in Ecolab, as really the
persistent state (which includes the stable state as a special case)
is the attractor. However, the inverse relationship between diversity
and connectivity does hold\cit{Standish98c}{(Standish, 1998)}, for
spatially dependent as well as panmictic cases. Therefore, in order
for diversity to show an increasing trend, a corresponding decreasing
trend must occur in connectivity. This ought to be true of the
biosphere also, given the universality of this relationship.

As mentioned in section \ref{Ecolab-sect}, the mutation operator is highly
conservative with respect to connectivity. It assumes that a new
species inherits the same connections as its parent, with random
additions or deletions according to a symmetric distribution (just as
likely to gain a connection as lose one). This has the effect of
preserving the connectivity over time. In order for connectivity to
decrease, different dynamics would need to be proposed, for example
assuming that the mutant species did not compete with its parent.

One possibility for the cause of this growth in diversity
is the mass extinctions, that have occurred a handful of times
throughout the Phanerozoic. However, the only reasonable way of
modeling this is to remove a random proportion of species from the
ecology at a particular time. This operation does not alter the
connectivity, as the links lost is exactly balanced by the reduced
diversity. When implemented within Ecolab, one gets the characteristic
rebound in diversity after the extinction event, however, the rebound
is back to about the same diversity level as existed prior to the extinction.

Another possibility that actually would work in the right way is
related to the fact that the Phanerozoic era corresponds to the
breakup of the Pangaea supercontinent --- firstly into Gondwana and
Laurasia, then into the six continents we know today. Assuming that
there is almost no migration between the continents (thus 6
equal-sized continents would support 6 times the diversity of one
continent that size) and that the species-area law within a continent
has $D\propto A^{.3}$, we would expect that a breakup of a single
supercontinent into 6 equal sized pieces should produce $6^{1-.3}=3.5$
times the diversity of the original supercontinent. This factor
accounts for a significant fraction of the diversity growth since the
Permian.\footnote{In case anyone thinks that this result is an
  argument in favour of habitat fragmentation for promotion of
  diversity, this is a question of scale. Over short timescales
  habitat fragmentation is bad for diversity, as is any major
  environmental change. Only over evolutionary timescales will the
  diversity bounce back.}\cit{Benton95}{(Benton, 1995)}

Clearly this is a very rough ``back of the envelope'' calculation, but
it is sufficient to show that continental breakup needs to be allowed
for in determining if there is any intrinsic evolutionary processes
driving diversity growth.

\section{Acknowledgements}

I wish to thank Mark Bedau for many illuminating discussions, and for
assistance in developing the neutral shadow model for Ecolab. I also
wish to thank the {\em New South Wales Centre for Parallel Computing}
for computational resources required for this project.

\bibliographystyle{plain}
\bibliography{rus}

\end{document}